\documentstyle[epsfig,12pt]{elsart}
\begin{document}
\newcommand{\p}{\partial}
\newcommand{\ls}{\left(}
\newcommand{\rs}{\right)}
\newcommand{\beq}{\begin{equation}} 
\newcommand{\eeq}{\end{equation}}
\newcommand{\beqa}{\begin{eqnarray}}
\newcommand{\eeqa}{\end{eqnarray}}
\newcommand{\bdm}{\begin{displaymath}}
\newcommand{\edm}{\end{displaymath}}
\newcommand{\ena}{\mbox{{\small 1}\hspace{-0.3em}1}}
\begin{frontmatter}
\title{Aspects of particle production in isospin asymmetric matter}
\author{G. Ferini},
\author{M. Colonna},
\author{T. Gaitanos},
\author{M. Di Toro\thanksref{dit}}
\address{Laboratori Nazionali del Sud INFN, I-95123 Catania, Italy,\\
and Physics-Astronomy Dept., University of Catania  }
\thanks[dit]{ditoro@lns.infn.it} 
\begin{abstract}
The production/absorption rate of particles in compressed and heated 
asymmetric matter is studied using a Relativistic Mean Field ($RMF$)
transport model with an isospin dependent collision term.
Just from energy conservation in the elementary production/absorption
processes we expect to see a strong dependence of the yields on the
basic Lorentz structure of the isovector effective interaction, 
due to isospin effects on the scalar and vector self-energies of the hadrons.
This will be particularly evident for the ratio of the rates of particles
produced with different charges: results are shown for 
$\pi^+/\pi^-$, $K^+/K^0$ yields.

In order to simplify the analysis we perform $RMF$ cascade simulations 
in a box with periodic boundary
conditions. In this way we can better pin down all such fine relativistic
effects in particle production, that could likely show up even
in realistic heavy ion collisions. In fact the box properties are
tuned in order to reproduce the heated dense matter formed during
a nucleus-nucleus collision in the few $AGeV$ beam energy region.

In particular, $K^{+,0}$  production is expected to be directly related to the
high density behaviour of the symmetry energy, since kaons are produced 
very early during the high density stage of the collision and their
mean free path is rather large.
We show that the $K^+/K^0$ ratio reflects important isospin contributions 
on the production rates just because of the large sensitivity around the 
threshold.
The results are very promising for the possibility of a direct link
between particle production data in exotic Heavy Ion Collisions ($HIC$)
and the isospin dependent part of the
Equation of State ($EoS$) at high baryon densities. 
\end{abstract}
\end{frontmatter}
\section{Introduction}
\label{chap1}

Recently the development of new heavy ion facilities (radioactive beams) 
has driven 
the interest on the dynamical behaviour of asymmetric matter, see the
recent reviews \cite{bao,baranPR}. 
Indeed, the isospin-dependent part of the
$EoS$ is of crucial 
importance in extrapolating structure calculations beyond the valley of 
stability 
and in astrophysical processes, such as structure and cooling of neutron
stars
 and supernovae 
explosions. 

Here we focus our attention on relativistic heavy ion collisions, that
provide a unique terrestrial opportunity to probe the in-medium nuclear
interaction in high density and high momentum regions. We will show that
particle production/absorption (here pions and kaons) processes in a dense 
and hot neutron rich
medium are nicely sensitive to the details of the relativistic structure of
the effective interaction. The results appear very promising for the 
possibility of directly pin down from data such fundamental microscopic
information.

Within a covariant picture of the nuclear mean field, 
for the study of asymmetric nuclear matter certainly the same 
considerations hold for the density behaviour of the symmetry term as in 
the iso-scalar case. The equilibrium conditions are directly related to the 
Lorenz structure of 
the iso-vector effective fields in a similar way as in the iso-scalar case 
(competition effects between attractive scalar and repulsive vector fields). 
However, for the description of the $a_{4}$ parameter of the Weizs\"{a}ecker 
mass formula (in a sense equivalent to the $a_{1}$ parameter for the 
iso-scalar part), extracted in the range from 28 to 36 MeV, there are 
different 
possibilities: (a) considering only the Lorentz vector $\rho$ mesonic field, 
and (b) both, the Lorentz vector $\rho$ (repulsive) and Lorentz scalar 
$\delta$ (attractive) effective 
fields \cite{kubis,liu}. The second assumption corresponds to the two 
strong effective $\sigma$ (attractive) and $\omega$ (repulsive) mesonic 
fields of the iso-scalar sector. 
In both cases one can fix the empirical value of the $a_{4}$ parameter, 
however important effects at supra-normal densities appear due to the 
introduction of the effective $\delta$ field. In fact 
the presence of an isovector scalar field is increasing the repulsive 
$\rho$-meson contribution at high baryon densities \cite{liu,baranPR}. 
This is a pure relativistic effect, due to the different Lorentz 
properties of these fields 
(the vector $\rho$ field grows with baryon density whereas the scalar 
$\delta$ field is supressed by the scalar density). 
 As a consequence, the isospin-dependent part of the
$EoS$ 
becomes ``stiffer'' when both fields are accounted for. This important 
feature appears in several models, see ref. \cite{gait04} where the 
covariant structure 
of the symmetry energy was investigated adopting different approaches to 
asymmetric nuclear matter such as Non-Linear $RMF$, 
Density Dependent Hadronic ($DDH$) and Dirac-Brueckner-Hartree-Fock ($DBHF$) 
theories. 
Moreover, relativistic approaches to asymmetric nuclear matter naturally 
lead to an effective ($Dirac$) mass splitting between protons and 
neutrons due to the 
Lorentz scalar nature of the $\delta$ field, \cite{kubis,liu,baranPR}. This 
is of relevance for the dynamics of heavy ion collisions and it has 
recently been 
microscopically studied within the $DBHF$ approach \cite{dbhf,diracmass}. 

The study of the influence of the high density symmetry energy on the 
$HIC$ dynamics has been recently started. 
So far one has considered collective flow observables, such 
as in-plane and elliptic flows \cite{dani,larionov,giessen,gait01} 
for protons and neutrons (or light isobars), 
isospin equilibration (or transparency) and pion production in heavy ion 
collisions at intermediate relativistic energies up to $1-2~AGeV$ 
\cite{bao2,gait04,gait03}. 
However, definite conclusions have not been drawn yet due to either the 
lack of experimental information on some observables
or the moderate dependence 
on the density behaviour of the symmetry energy. Comparisons with recent 
experimental data on isospin transparency 
seem to suggest a $stiff$ high density behaviour of the 
symmetry energy, as predicted by $RMF$ theory with the inclusion 
of the $\delta$ meson \cite{gait03,chen}.

In order to explore the symmetry energy at supra-normal densities one  
has to select signals directly emitted from the earlier non-equilibrium 
high density 
stage of the heavy ion collision, possibly without any secondary interaction 
with the hadronic 
environment.

In fact particle production is of particular interest since it mainly 
takes place in the first binary 
collisions during the formation of the high density phase, where the energy 
densities are still above resonance thresholds. 
 This has stimulated recent
experimental studies
\cite{pions,kaos}. However, the 
investigation of particle production represents  
a non-trivial task due to the following reasons: 
\begin{itemize}
\item {(a) At intermediate 
energies ($<~1.6~ AGeV$) some of the pions created 
during the high density phase are directly emitted, but many of them undergo 
secondary re-scattering with isospin exchange before emission, \cite{gait04}.}
\item{(b) Production and reabsorption processes must be consistently
 implemented in relativistic transport models with isospin effects. In fact  
within an 
isospin dependent mean field, isospin exchange processes such as 
$R^{\prime} \leftrightarrow \pi N,~\pi\pi R$ ($N,R,R^{\prime}$ stand 
for protons/neutrons and the $\Delta^{\pm,0,++}$ isospin resonance states, 
respectively,  and $\pi^{\pm,0}$ for pions) modify the scalar and vector 
content of the available energy between the in- and outgoing channels. This 
has to be taken into account in the energy conservation, and it is
particularly relevant around the thresholds.} 
\item{(c) The situation for the production of particles with 
strangeness such as kaons ($K^{\pm,0}$) is similar, but even more complicated 
due to the availability of more than $50$ isospin channels for the creation 
of $K^{+}$ and $K^{0}$, only from $BB$ and $B\pi$ processes ($B$ stands for 
a resonance or a nucleon).} 
\end{itemize}

Kaons 
are produced very early during the high density stage of the collision,
see the compressibility dependence in refs. \cite{kaons,fuchshyp}.
Moreover, the production of $K^{+,0}$ is expected to be 
directly 
related to the high density behaviour of the symmetry energy, since the mean 
free path of these low energy
kaons in nuclear matter  
is rather large ($\lambda(K^{+,0}) \sim 7$ fm \cite{kaons,fuchshyp}). 
These two facts make the subthreshold kaon production a very 
promising probe for the high density symmetry energy. Anti-kaons ($K^{-}$) 
do not fulfill the desired criteria due to their relatively small 
mean free path value (strong reabsorption effects).
A related problem for $K^{-}$'s is the importance of off-shell 
contributions in the propagation and the corresponding uncertainty on the
self-energy evaluations \cite{CassingNPA727}. At variance the quasiparticle 
approach followed here in a transport $RMF$ frame is consistently justified
for $K^{+,0}$ mesons. In any case the anti-kaon production is not so relevant 
at the energies of interest in the present study 
\cite{WeberPRC67,ReiterNPA722}.  

In this article we investigate {\it hadronic matter} properties in a 
finite box, with periodic boundary conditions, performing cascade calculations 
by an extended isospin dependent collision integral.
The temperature and baryon density conditions are tuned in order to
reproduce the early stage of a realistic heavy ion collision at beam
energies around the threshold for kaon production. Varying the asymmetry
of the system 
we can then study the influence of the isovector part of the hadronic
in medium interaction.

Thus,  we 
discuss particle production in fixed conditions of density and temperature 
(for both symmetric and asymmetric matter). In this way we can have a 
precise control on the production/absorption rates in the different
channels.
 
The description of the mean field is important, since nucleons and resonances
are {\it dressed} by the mean field self-energies. This  will
directly affect the threshold energies, in particular for inelastic channels. 
We first give an outline of the mean field models of asymmetric 
nuclear matter, before passing to the description of the box 
calculations (Sect. \ref{chap2}). The modification of the threshold 
conditions according to energy-momentum conservation is a 
non-trivial task which 
will be discussed in Sect. \ref{chap3}. 
The isospin dependence of the chemical equilibrium conditions, to which
particle multiplicities obey after a given time interval $t_{eq}$, is
illustrated in Sect. \ref{new}, considering in particular the 
$\pi^-/\pi^+$ ratio. 
The results are 
presented in Sect. \ref{chap4}. We will first discuss isospin effects 
on resonances and pion production, before passing to kaons. The results 
provide an important guide to the understanding of particle production in 
heavy ion collisions at relativistic energies.

\section{Cascade calculations for hadronic asymmetric matter}
\label{chap2}
\subsection{Relativistic effects on the symmetry energy}

\nopagebreak
\begin{table}[t]
\begin{center}
\begin{tabular}{|l|c|c|c|c|c|c|c|c|c|}
\hline\hline 
       & $f_{\sigma}$ ($fm^2$)   & $f_{\omega}$ ($fm^2$) & $f_{\rho}$ ($fm^2$)
     & $f_{\delta}$ ($fm^2$)     & A ($fm^{-1}$) & B &  \\ 
\hline\hline
   $NL$          &     9.3        &     3.6     & 0.0 &    0.0     &
  0.015     &    -0.004  \\ 
\hline
   $NL\rho$          &     9.3        &     3.6     & 1.22 &    0.0     &
  0.015     &    -0.004  \\ 
\hline
   $NL\rho\delta$ &     9.3        &     3.6    & 3.4 &    2.4     &
  0.015     &    -0.004  \\ 
\hline\hline
\end{tabular}
\end{center}
\vskip 0.5cm
\caption{\label{table1} 
Coupling parameters in terms of $f_{i} \equiv 
(\frac{g_{i}}{m_{i}})^{2}$ 
for $i=\sigma,~\omega$, $f_{i} \equiv 
(\frac{g_{i}}{2m_{i}})^{2}$ for $i=\rho,~\delta$, 
$A \equiv \frac{a}{g_{\sigma}^{3}}$ and 
$B \equiv \frac{b}{g_{\sigma}^{4}}$ for the non-linear $NL$ models \cite{nl} 
using the $\rho$ ($NL\rho$) and 
both, the $\rho$ and $\delta$ mesons ($NL\rho\delta$) for the 
description of the isovector mean field.
The $NL$ model does not contain any isospin dependence.}
\vskip 0.5cm
\end{table}
Within a relativistic framework the energy conservation in binary 
collisions between hadrons in nuclear matter
is directly related to the Lorentz components of the 
hadron self-energy $\Sigma=\Sigma_{s} - \gamma^{\mu}\Sigma_{\mu}$. 
The scalar and vector self-energies $\Sigma_{s},~\Sigma^{\mu}$ generally 
depend on baryon density and momentum, according to microscopic $DBHF$ 
calculations \cite{dbhf2}. However, since we are focusing here on 
isospin effects, 
i.e. on the iso-vector part of the self-energies, and $DBHF$ studies of 
asymmetric 
nuclear matter are still rare or just being started \cite{dbhf,diracmass}, 
we will use 
a simple phenomenological version of the Non-Linear (with respect to the 
iso-scalar, Lorentz scalar $\sigma$ field \cite{nl}) Walecka model 
which corresponds 
to the 
Hartree or Relativistic Mean Field ($RMF$) approximation within the 
Quantum-Hadro-Dynamics \cite{qhd}. 
According to this model the baryons (protons and neutrons) are described by an 
effective Dirac equation $(\gamma_{\mu}k^{*\mu}-m^{*})\Psi(x)=0$, whereas the 
mesons, which generate the classical mean field, are characterized by 
corresponding covariant equations of motion in the Local Density 
Approximation, with coupling constants that are not density dependent. 
The presence of the hadronic medium modifies the masses and momenta 
of the hadrons, i.e. $m^{*}=M+\Sigma_{s}$ (effective masses), 
 $k^{*\mu}=k^{\mu}-\Sigma^{\mu}$ (kinetic momenta). 
For asymmetric matter the self-energies are different for protons and 
neutrons. This depends on the way the iso-vector mean field is described. 
As discussed in the introduction, in contrast to the iso-scalar case, since 
here one does not have the stringent saturation condition of balancing 
attractive 
(scalar) and repulsive (vector) contributions, one can apply either only the 
$\rho$ meson, or both, the $\rho$ and $\delta$ mesons. We will call the 
corresponding models as $NL\rho$ and $NL\rho\delta$ models, respectively. 
Furthermore, 
we will also use the non-linear Walecka model without accounting for isospin 
dependent mean field ($NL$-model). 

The parameters used here are summarized in Table \ref{table1}.
The corresponding saturation properties of symmetric nuclear matter
are $E/A(MeV)=-16.0$, $\rho_0(fm^{-3})=0.16$, $K(MeV)=240$, $m^*/M=0.75$
and the symmetry energy parameter (for $NL\rho, NL\rho\delta$) is 
$a_4(MeV)=30.7$
 \cite{liu,baranPR}.
These Lagrangians have been already used for flow \cite{greco}, pion production
 \cite{gait04} and isospin tracer \cite{gait03} calculations in relativistic
$HIC$ with an overall good agreement to data. Moreover it has been recently 
shown that at high baryon densities the $EoS$ of more microscopic
$DBHF$ calculations can be well reproduced \cite{gait04,liuns}.

In particular, for the more general $NL\rho\delta$ case one obtains for 
the self-energies 
of protons and neutrons:
\begin{eqnarray}
\Sigma_{s}(p,n) & = & - f_{\sigma}\sigma(\rho_{s}) \pm f_{\delta}\rho_{s3} 
\label{sigs}\\
\Sigma^{\mu}(p,n) & = & f_{\omega}j^{\mu} \mp f_{\rho}j^{\mu}_{3},
\label{sigv}
\quad .
\end{eqnarray}
(upper signs for neutrons), 
where $\rho_{s}=\rho_{sp}+\rho_{sn},~
j^{\alpha}=j^{\alpha}_{p}+j^{\alpha}_{n},\rho_{s3}=\rho_{sp}-\rho_{sn},
~j^{\alpha}_{3}=j^{\alpha}_{p}-j^{\alpha}_{n}$ are the total and 
isospin scalar 
densities and currents and $f_{\sigma,\omega,\rho,\delta}$  are the coupling 
constants of the various 
mesonic fields, see Table \ref{table1}. 
$\sigma(\rho_{s})$ is the solution of the non linear 
equation for the $\sigma$ field \cite{liu,baranPR}.

\begin{figure}[t]
\unitlength1cm
\begin{picture}(10.,9.0)
\put(0.0,0.0){\makebox{\epsfig{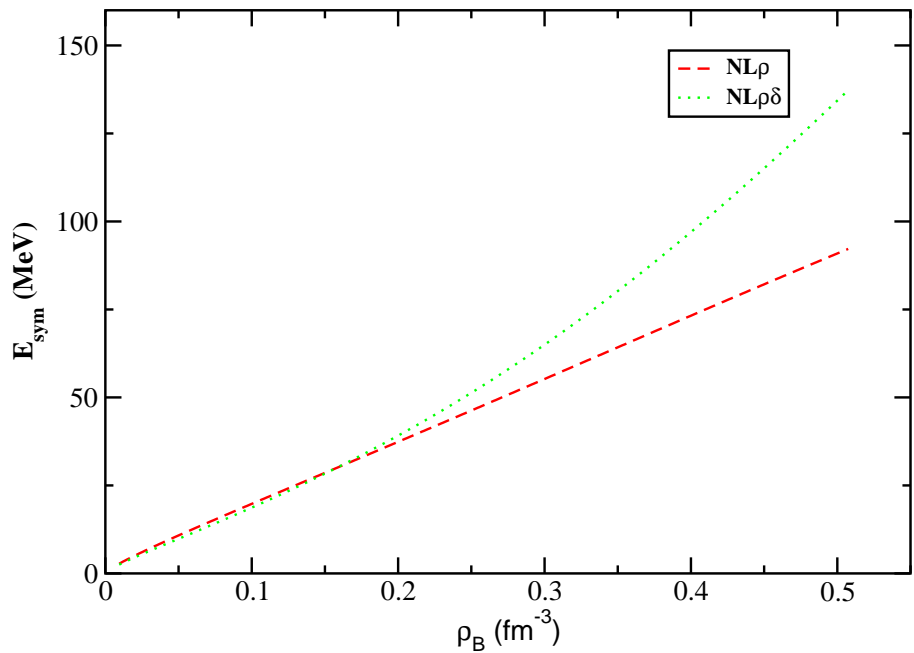}}}
\end{picture}
\caption{Density dependence of the symmetry energy 
 Eq.(\protect\ref{esym3}) including only the $\rho$-meson (dashed line, 
$NL\rho$-model) and both, the $\rho$ and $\delta$ mesons (dotted line, 
$NL\rho\delta$-model) in the description of asymmetric matter.
}
\label{fig1}
\end{figure}

Within the relativistic description we expect to see a series of isospin 
effects on the reaction dynamics just due to the changes in 
 the covariant Lorentz structure of 
the isovector interaction.
These different isospin contributions of relativistic origin can be 
already seen in the 
expression for the symmetry energy \cite{liu,baranPR}: 
\begin{equation}
E_{sym} = \frac{1}{6} \frac{k_{F}^{2}}{E_{F}^*} + 
\frac{1}{2}
\left[ f_{\rho} - f_{\delta}\left( \frac{m^{*}}{E_{F}^*} \right)^{2}
\right] \rho_{B}
\label{esym3}
\quad ,
\end{equation}
with $E_{F}^* \equiv \sqrt{k_{F}^{2} + {m^*}^2}$.
When including the scalar iso-vector $\delta$-meson the 
asymmetry interaction parameter is given by the combination 
$[ f_{\rho} - f_{\delta}( \frac{m^{*}}{E_{F}^*})^{2}]$ of the 
repulsive vector ($\rho$) and attractive ($\delta$) iso-vector couplings. 
Thus, in order 
to reproduce the same bulk asymmetry parameter $a_{4}$, we  have to increase 
the $\rho$-meson coupling when the 
$\delta$-meson is considered in the iso-vector part of the equation of state.
 The net effect is 
a stiffer symmetry energy at higher baryon densities, due to the
$\frac{m^{*}}{E_{F}^*}$ quenching of the attractive part,
as shown in 
Fig. \ref{fig1}. This is a general feature 
that has been seen to be important in the 
dynamical 
simulations \cite{gait04}. 
Other relativistic effects of the Lorentz structure of the isovector
interaction, particularly important for the inelastic channels,
will be:
\begin{itemize}
\item{
The vector field $\Sigma^{\mu}$, Eq.(\ref{sigv}), will show a 
different isospin dependence 
since in the $NL\rho\delta$ model the $\rho$-meson coupling is larger with 
respect to that of the $NL\rho$ model. In order to conserve the total energy 
in 
inelastic collisions we have to include the isospin dependence of the 
vector fields between the ingoing and outgoing 
channels, as it will be  discussed in detail in the next section.
Thus, the particle production/absorption will be directly sensitive
to the covariant form of the isovector interaction.} 
\item{
Another $\delta$-effect, of interest for transport properties, will be
the splitting of the neutron/proton effective masses \cite{liu,greco}, 
see Eq.(\ref{effmass}), which will be also discussed later on in detail 
in terms of the energy for subthreshold  
particle production: }
\end{itemize}
\begin{equation}
m^{*}_{i}=M+\Sigma_s(i)=M - f_{\sigma}\sigma(\rho_s) \pm f_{\delta}\rho_{s3}
\quad\mbox{(+ n, - p)}\quad
\label{effmass}
\quad .
\end{equation}

\subsection{Hadronic matter in a box}

In order to study hadronic matter, we confine the particles in a 
cubic box with periodic boundary conditions. The number of nucleons and the
box dimensions are given as inputs of the simulation and can thus be changed
to explore different values for the baryon density $\rho_B$ and 
the asymmetry parameter $\alpha$ of the system. 
We have fixed the total number of 
nucleons to {\it A}=100, varying the relative proton and neutron numbers
 {\it $Z$} and {\it $N$} in order to get 
$\alpha=\frac{N - Z}{N + Z}=
\frac{\rho_{Bn}-\rho_{Bp}}{\rho_{Bn}+\rho_{Bp}}$
in the range from 0.0 (symmetric nuclear matter) to 0.4 (which corresponds to 
extremely asymmetric matter, with $N/Z=2.3$). The box dimensions are set
in order to achieve a baryon density (kept constant during the temporal 
evolution of the system) of roughly 2.5 times the saturation density, 
while for the temperature the value of 
T=60 MeV has been chosen. These parameters should resemble the conditions 
achieved during the dense 
phase in heavy ion collisions at few $AGeV$ energies \cite{essler}. 

\begin{figure}[t]
\unitlength1cm
\begin{picture}(10.,9.0)
\put(0.0,0.0){\makebox{\epsfig{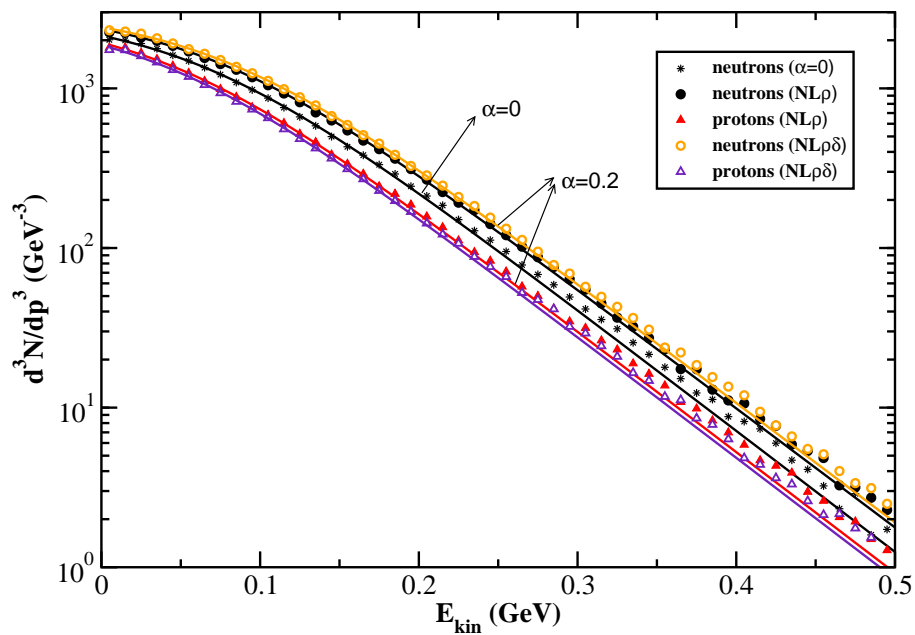}}}
\end{picture}

\caption{Kinetic energy spectra $(E_{kin} \equiv E_i^*-m_i^*, {\bf p} \equiv
 \hbar {\bf k}^*)$ of nucleons for symmetric and asymmetric 
nuclear 
matter at $t=0$ fm/c and fixed baryon density $\rho_{B}=2.5~\rho_0$. 
stars: neutron spectra for symmetric matter 
($\alpha=0$); filled circles and triangles: neutron and proton kinetic energy 
spectra for asymmetric matter ($\alpha=0.2$) in the $NL\rho$ model; 
open circles and triangles: neutron and proton kinetic energy 
spectra for asymmetric matter ($\alpha=0.2$) in the $NL\rho\delta$ model. 
The corresponding solid lines are least-square fits according to a Fermi-Dirac 
distribution at the input temperature of $T=60$ MeV. 
}
\label{fig2}
\end{figure}
At t=0 fm/c neutrons and protons are randomly distributed in
the box volume and their kinetic $3$-momenta  ${\bf k}^*$ are generated 
according to relativistic Fermi-Dirac distribution functions at the given
 temperature T:
\begin{equation}
n_{i}({\bf k}^{*}) = \frac{1}{1+exp[(E_{i}^{*}({\bf k}^{*})-\mu_{i}^{*})/T]} 
\label{fermi}
\quad .
\end{equation}

In eq. (\ref{fermi}) $n_i$ (i={\it n},{\it p}) are the occupation numbers 
for neutrons
and protons, $E_{i}^{*}=\sqrt{{\bf k}_{i}^{*2}+m_{i}^{*2}}$,
while $\mu_{i}^{*}$ (i=p,n) are the effective chemical potentials for protons 
and neutrons respectively, given by ($\epsilon$ is the energy density):
\begin{equation}
\frac{\partial{\epsilon}}{\partial{\rho}_{i}} = \mu_{i} = \mu_{i}^{*}+
f_{\omega}\rho_{B}\mp f_{\rho}\rho_{B3}~~~~~(-~n, +~p). 
\label{chempot}
\end{equation}
At zero temperature the $\mu_{i}^{*}$ reduce to $E_{Fi}^*$ (i=p,n).

Once both T and $\alpha$ (and hence $\rho_{Bi}$) are fixed, $\mu_{i}$, the 
scalar densities $\rho_{si}$ and effective masses $m_{i}^{*}$ can be 
calculated 
self-consistently solving the system of equations
\begin{eqnarray}
\rho_{Bi} & = & \frac{2}{(2\pi)^{3}} \int d^{3}{\bf k}^* n_{i}({\bf k}^{*}) 
\label{rhob}\\
m_{i}^{*} & = & M - f_{\sigma}\sigma (\rho_{s})\pm f_{\delta}(\rho_{sp}-
\rho_{sn}) 
\quad\mbox{(+ n, - p)}\quad
\label{emstar}\\
\rho_{si} & = & \frac{2}{(2\pi)^{3}} \int d^{3}{\bf k}^* \frac{m_{i}^{*}}
{E_{i}^{*}}
  n_{i}({\bf k}^{*}). 
\label{rhos}
\end{eqnarray}

For symmetric nuclear matter 
($\alpha$=0)  $\rho_{Bi}$, $\rho_{si}$, $\mu_{i}$ and $m_{i}^{*}$ are the same 
for both neutrons and protons 
so that there is no difference in their momentum distributions. In the case of 
asymmetric matter the momentum distributions of neutrons and protons are 
slightly different 
due to the different radii of the respective Fermi spheres so that
 $\mu_{p}^{*}\neq\mu_{n}^{*}$;
furthermore, the inclusion of isovector scalar meson $\delta$ implies 
$m_{p}^{*}\neq m_{n}^{*}$. Anyway the initial effective momentum 
distributions of 
both protons
and neutrons just sligthly differ in $NL$, $NL\rho$ and $NL\rho\delta$ 
models.  
This can be seen in Fig. \ref{fig2}, where the nucleon spectra for 
$\alpha$=0 and 
$\alpha$=0.2 are shown.

We then let the particles (nucleons, $\Delta$ resonances and pions) propagate 
according to a free kinetic equation plus the collision term, in the spirit
of a cascade calculation. The mean-field potentials, constant once 
($\rho_B, \alpha$) are fixed, are accounted for in the energy-momentum
conservation for the various channels of the collision term.
 The boundary conditions are modeled in such a way that 
particles escaping from the box are reinserted at the opposite
side with the same momentum. In their motion nucleons undergo 
elastic and inelastic 
binary collisions, thus 
exciting resonances that further decay into pions, and then producing strange 
particles such as hyperons and kaons. Here we consider the lowest mass 
resonances only, namely $\Delta$(1232). 

Because of inelastic collisions, a small fraction of the 
initial kinetic energy of 
the system is lost in particle production and after 
150 fm/c nucleons momenta are 
distributed according to a relativistic Fermi-Dirac function 
with temperature T=56 MeV.
Pion spectra do not appear thermalized yet, though pion multiplicities
saturate already after $t \approx 30~fm/c$. 
Indeed longer time scale are needed to reach full thermal 
equilibrium \cite{bratko}. 

For elastic nucleon-nucleon cross sections we use the free 
parametrizations of Cugnon et al. \cite{cugnon}, in order to
reduce the computing, \cite{sigma}. 
For the inelastic channels (resonance production) we follow 
the cross sections of Ref. \cite{huber}. 
Kaon production is treated perturbatively, so that it does not 
affect the overall dynamics. It occurs according to two different creation 
mechanisms, i.e. baryon induced processes such as BB$\rightarrow$BYK, and 
pion induced processes such as $\pi$B$\rightarrow$YK. In both cases, kaons 
are created together with $Y$ hyperons ($\Lambda$ and $\Sigma$) 
to ensure 
strangeness conservation and are 
not reabsorbed by nucleons. They can anyway undergo elastic rescattering 
with nucleons, and thus modify their momentum distribution. However, 
since we are just interested in kaon yields and their possible 
dependence on the
 symmetry energy, we neglect this effect here and do not propagate 
kaons after production. 
The elementary cross sections
for kaon production from $BB$ and $\pi B$ channels are taken from 
Refs. \cite{tsu1} and \cite{tsu2}, respectively. 
To treat all these scattering processes in infinite hadronic asymmetric matter 
we have modified the original version of the collision term of the T\"{u}bingen
$RQMD$ simulation code \cite{fuchshyp}, taking explicitly into account 
the isospin 
dependence of  scalar and vector self-energies $\Sigma_{s}, \Sigma^{\mu}$. 
This turns into an isospin dependence of the threshold of all isospin channels 
and thus affects particle yields, in a way that will be  discussed in 
detail in the following section.

\section{Isospin dependent threshold conditions}
\label{chap3}

The crucial requirement of two-body collisions is the energy-momentum 
conservation, i.e. the conservation of the total invariant collision 
energy $s$, which, in terms of ingoing (1,2) and
outgoing (3,4) canonical momenta ($k_{1,2}^{\mu}$, $k_{3,4}^{\mu}$),
reads: 
\begin{equation}
s_{in} = (k_{1}^{\mu}+k_{2}^{\mu})(k_{1\mu}+k_{2\mu})
           = (k_{3}^{\mu}+k_{4}^{\mu})(k_{3\mu}+k_{4\mu})
           = s_{out}
\label{scan}
\qquad .
\end{equation}

If the isospin degree of freedom is not accounted for in inelastic collisions,
  we can simply fulfill 
energy-momentum conservation using kinetic 4-momenta $k_i^{*\mu}$ and the
corresponding invariant collision energy
\begin{equation}
s^{*}_{in} = (k_{1}^{*\mu}+k_{2}^{*\mu})(k_{1\mu}^{*}+k_{2\mu}^{*})
           = (k_{3}^{*\mu}+k_{4}^{*\mu})(k_{3\mu}^{*}+k_{4\mu}^{*})
           = s^{*}_{out}
\label{skin}
\qquad .
\end{equation}

The relation between kinetic and canonical momenta is 
$k^{*\mu}_{i} \equiv k^{\mu}_{i}-\Sigma^{\mu}_{i},~(i=1,2,3,4)$ and the 
effective masses $m^{*}_{i}=M+\Sigma_{si}$ enter in the energy conservation 
via the $0$-component of the kinetic $4$-momentum 
$E^{*}=\sqrt{m^{*2}+k^{*2}}$. 

The condition (\ref{skin}) is just 
a constraint on the kinetic plus the effective mass energy of 
the colliding 
particles and it is sufficient if one considers symmetric ($N=Z$) matter 
or restricts itself to elastic collisions or just to the
cases where the mean field 
(scalar and vector self-energies) does not change between the 
ingoing and the
outgoing particles. In those particular cases 
Eq. (\ref{skin}) implies Eq. (\ref{scan}). 

In the general case of asymmetric nuclear matter, however, the introduction of 
the iso-vector effective mesons causes a splitting in the scalar 
{\it and} in the vector fields between protons and neutrons, as discussed 
in the 
previous sections (see Eqs. (\ref{sigs},\ref{sigv})). For the other hadrons 
the following expressions in terms of the nucleon self-energies will be used. 
For the four isospin states of the $\Delta$ resonance one has 
\cite{bao02} ($i=scalar,vector$)
\begin{eqnarray}
\Sigma_{i}(\Delta^{-}) & = & \Sigma_{i}(n)
\label{sigma_dm}\\
\Sigma_{i}(\Delta^{0}) & = & \frac{2}{3}\Sigma_{i}(n)+\frac{1}{3}\Sigma_{i}(p)
\label{sigma_d0}\\
\Sigma_{i}(\Delta^{+}) & = & \frac{1}{3}\Sigma_{i}(n)+\frac{2}{3}\Sigma_{i}(p)
\label{sigma_dp}\\
\Sigma_{i}(\Delta^{++}) & = & \Sigma_{i}(p)
\label{sigma_dpp}
\quad ,
\end{eqnarray}
in which the weights (1, $\frac{1}{3}$, $\frac{2}{3}$) or 
`isospin coefficients' are just 
the square 
of the Clebsch-Gordon coefficients for isospin coupling in the 
$\Delta \Longleftrightarrow \pi N$ processes. 
In the same framework, for hyperons $\Lambda,\Sigma^{0,\pm}$ we have:
\begin{eqnarray}
\Sigma_{i}(\Lambda) & = & \frac{2}{3}
\left( \frac{\Sigma_{i}(p)+\Sigma_{i}(n)}{2} \right)
\label{hyp_l}\\
\Sigma_{i}(\Sigma^{0}) & = & \frac{2}{3}
\left( \frac{\Sigma_{i}(p)+\Sigma_{i}(n)}{2} \right)
\label{hyp_s0}\\
\Sigma_{i}(\Sigma^{+}) & = & \frac{2}{3}\Sigma_{i}(p)
\label{hyp_sp}\\
\Sigma_{i}(\Sigma^{-}) & = & \frac{2}{3}\Sigma_{i}(n)
\label{hyp_sm}
\quad ,
\end{eqnarray}
where the hyperon fields are further scaled by a factor  
$\frac{2}{3}$ \cite{fuchshyp},
since, according to the quark model, there are only two light quarks in a 
hyperon instead of the three light quarks in a nucleon.
Pions and kaons are assumed to be unaffected by the nuclear mean field 
and then have 
zero self-energies. This assumption is not  strictly correct, at least for 
kaons, which 
should be sensitive to medium effects and thus shift their 
production thresholds. 
Anyway, such a shift produced by a kaon potential should be almost the same 
for both 
$K^{0,+}$ \cite{kaons} and thus it should not significantly affect the 
$K^{+}/K^{0}$ ratio we will discuss here. For this 
reason  we do not take such a potential into account in the present work.

In inelastic collisions, due to isospin exchange, the scalar and vector 
self-energies 
between the ingoing and outgoing channels may differ, so that 
Eq. (\ref{skin}) does 
no longer imply Eq. (\ref{scan}). A typical example is the inelastic process 
$nn \longrightarrow p\Delta^{-}$ from which a $\Delta^{-}$ resonance 
is formed. 
It is clear that, even in the presence of the $\rho$ meson only, the vector 
self-energy 
of the final channel changes and in the general case of accounting for both, 
the $\rho$ and $\delta$ mesons, both self-energies (scalar and vector) 
of the outgoing channel differ from those of the ingoing channel. 

In order to properly impose energy-momentum conservation one thus needs 
to replace 
condition (\ref{skin}) with (\ref{scan}). It then follows from 
Eq. (\ref{scan}), 
after some algebra, that the effective momenta ${\bf k}^{*}_{3,4}$ of 
the outgoing 
particles (in the $cm$ local reference frame in which 
${\bf k}^{*}_{1}+{\bf k}^{*}_{2}={\bf k}^{*}_{3}+{\bf k}^{*}_{4}={\bf 0},
~|{\bf k}^{*}_{out}| \equiv |{\bf k}^{*}_{3}| \equiv |{\bf k}^{*}_{4}| $) are 
linked to the total energy $s_{in}$ before the collision through the expression
\begin{equation}
\underbrace{-(\Sigma^{0}_{3}+\Sigma^{0}_{4}) + 
\sqrt{s_{in}+({\bf \Sigma}_{3}+{\bf \Sigma}_{4})^{2}}}_{\tilde{s}}
 = \sqrt{m^{*2}_{3}+k^{*2}_{out}} + \sqrt{m^{*2}_{4}+k^{*2}_{out}}
\label{smod}
\quad .
\end{equation}
It is worthwhile to stress that in the limiting case of no isospin exchange 
(elastic processes) or isospin independent self-energies (no isospin 
splittings 
in the scalar and vector self-energies) the general condition (\ref{smod}) 
reduces to the usual energy conservation relation, i.e.
 $\tilde{s} \longrightarrow \sqrt{s^{*}}$ 
with $s^{*}$ given by Eq. (\ref{skin}).
Eq. (\ref{smod}) implies the following threshold condition for a given 
inelastic process:
\begin{equation} 
s_{in}\geq \underbrace {(m_{3}^{*}+\Sigma _{3}^{0}+m_{4}^{*}+
\Sigma _{4}^{0})^2 - 
({\bf \Sigma}_{3}+{\bf \Sigma}_{4})^2}_{s_{0}}
\label{thresh}
\quad .
\end{equation}
This condition reduces to $s_{in}^{*}\geq (m_{3}^{*}+m_{4}^{*})^2$ whenever 
we have no isospin
 dependence, i.e. for symmetric nuclear matter and in general for the 
$NL$ model and for elastic 
collisions. The multiplicity of particles produced in the inelastic 
processes allowed by Eq.
(\ref{thresh}) depends on the available energy above the threshold, i.e. 
on the difference
$\Delta s \equiv  s_{in}-s_{0}$, which is affected by the isovector channel 
through both the effective masses
and the vector self-energies and therefore it changes when considering the 
$NL$, $NL\rho$ and
$NL\rho\delta$ models. 
It is appropriate to discuss such variations of $\Delta s$ for the 
cases of ${\it nn}$, 
${\it pp}$ and ${\it np}$ collisions. The $rhs$ of Eq. (\ref{thresh}) is not 
or slighlty affected 
by both isovector scalar ($\Sigma_{s}$) and vector ($\Sigma^{0}$) mesons 
depending 
on the isospin state of the outgoing channel. In the particular case of a  
process $nn \longrightarrow p\Delta^{-}$ ($pp \longrightarrow n\Delta^{++}$)
 $s_{0}$ is 
respectively given by:
\begin{eqnarray}
s_{0} & = & 
\left[ m_{p} + \Sigma_{s}(p) + \Sigma^{0}(p) + m_{\Delta} + \Sigma_{s}
(\Delta^{-}) + 
\Sigma^{0}(\Delta^{-}) \right]^2 
\nonumber\\
& & -\left[ {\bf \Sigma}(p)+{\bf \Sigma}(\Delta^{-}) \right]^2
\nonumber\\
s_{0} & = &  
\left[ m_{n} + \Sigma_{s}(n) + \Sigma^{0}(n) + m_{\Delta} + \Sigma_{s}
(\Delta^{++}) + 
\Sigma^{0}(\Delta^{++}) \right]^2 
\nonumber\\
& & -\left[ {\bf \Sigma}(n)+{\bf \Sigma}(\Delta^{++}) \right]^2
\quad ,
\end{eqnarray}

from which it is clear that, according to Eqs.(\ref{sigs}-\ref{sigv},
\ref{sigma_dm}-\ref{sigma_dpp}),
cancellation occurs between $\Sigma_{i}(p)$ and $\Sigma_{i}(\Delta^{-})$. 
Hence it follows 
 that $s_{0}$
is the same in all the three discussed models. For the other 
inelastic $nn$ and $pp$ channels
 and in the case of $np$ inelastic collisions, 
which proceed through the excitation of the $\Delta^{0}$ or 
$\Delta^{+}$ state of the $\Delta$ 
resonance, this cancellation is only partial, but in any case the net 
contribution on 
$s_{0}$ is moderate. 
The main effects on $\Delta s \equiv s_{in}-s_{0}$ are therefore 
coming from $s_{in}$. 
According to Eq. (\ref{scan}) 
we have (always in the $cm$ local reference frame)
\begin{equation} 
s_{in} = 
\left( E^{*}_{1}+\Sigma^{0}_{1}+E^{*}_{2}+\Sigma^{0}_{2}
\right)^{2}
- \left( 
{\bf \Sigma}_{1}+{\bf \Sigma}_{2}
\right)^{2}
\label{sin}
\quad .
\end{equation}
Since we are dealing with infinite nuclear matter at rest, in which the 
spatial currents vanish, the last term of Eq. (\ref{sin}) is
almost negligible also in the local frame considered here. 
Furthermore, due to moderate variations of $E^{*}$ induced by $m^{*}$, 
$\Sigma^{0}$ gives the major contribution to $s_{in}$ variations
in the different interaction models. 

For $nn$, $pp$ and $np$ 
collisions  we have, respectively 
\begin{eqnarray}
s_{in} & = & 4  \left[ (E_{n}^{*} + \Sigma^{0}_{n})^2  \right] \nonumber
\\
s_{in} & = & 4  \left[ (E_{p}^{*} + \Sigma^{0}_{p})^2  \right] \nonumber
\\
s_{in} & = & (E_{n}^{*} + E_{p}^{*} + \Sigma^{0}_{n} +\Sigma^{0}_{p})^2  
\label{snp}
\quad .
\end{eqnarray}

In the $NL$ model $\Sigma^{0}$ is the same for neutrons and protons and 
hence $s_{in}$
is the same for all $NN$ collisions (for fixed momenta). 
When adding the isovector vector $\rho$-meson only ($NL\rho$ model), 
$\Sigma^{0}_{n}$ increases 
(see Eq. \ref{sigv}) with respect to the $NL$ case, while the opposite 
occurs for 
$\Sigma^{0}_{p}$.
This effect is enhanced in the case of $NL\rho\delta$ model, as the 
introduction 
of the $\delta$ increases the $\rho$-meson coupling by about a factor $3$.
Correspondingly, $s_{in}$
increases (decreases) for $nn$ ($pp$) channels when turning the 
isovector channel on, 
both  without and with the
isovector scalar meson. At variance, in the case of $np$ collisions
due to cancellation effects in the sum of $(n,p)$ vector self-energies, 
$s_{in}$ is exactly the same when the $\rho$-meson is included
and it will show very small variations passing from $NL\rho$ to 
$NL\rho\delta$ (due to the $(n,p)$ effective mass splitting).

All that turns out into
a gradual increasing (decreasing) multiplicity of 
$\Delta^{-}$ ($\Delta^{++}$) states
when going from $NL$ to $NL\rho$ and then to $NL\rho\delta$ model 
and hence affects
the relative populations of different isospin states for pions and 
kaons, as we will widely 
discuss in the Sect. \ref{chap4}.

\section{Isospin dependent equilibrium conditions}
\label{new}
Since calculations are performed within a box with boundary conditions
(particles cannot escape) and production and reabsorption processes
are consistently taken into account for $\Delta$'s and pions, 
we observe that, after a given time interval $t_{eq}$, 
chemical equilibrium is reached
among the different particle populations. 
Indeed, starting from $t_{eq}$, $\Delta$ and pion multiplicities do not
evolve anymore, indicating that the production rate equals the 
reabsorption rate. 
When chemical equilibrium is established, very simple relations hold
and the chemical potential of produced particles, as pions, 
can be expressed in terms of neutron and proton chemical potentials,
 $\mu_n$ and  $\mu_p$.
In particular, at equilibrium, the chemical potential of $\pi^-$ is
equal to the difference $\mu_n - \mu_p$, while the chemical 
potential of  $\pi^+$ is equal to  $\mu_p - \mu_n$. 
Hence one can already make analytical predictions for the 
$\pi^-/\pi^+$ ratio in the different models used here. 
This ratio can be expressed as:
\begin{equation}
 \pi^-/\pi^+ \propto exp~( 2( \mu_n - \mu_p )/ T), 
\label{pireq}
\end{equation}
where $T$ is the temperature of the system.
Neutron and proton chemical potentials have different values in the models
we consider ($NL$, $NL\rho$ or $NL\rho\delta$) because the self-energies
are different. 
In particular, going from $NL$ to  the $NL\rho$ model, we expect that the 
$\pi^-/\pi^+$ ratio increases by a factor $exp~(4f_\rho
(\rho_{Bn}-\rho_{Bp})/T)$,
according to the variation of the difference  $\mu_n - \mu_p$
(see Eq.(6)). 
A similar, though weaker, increase of the  $\pi^-/
\pi^+$
ratio is expected also   
when going from   $NL\rho$ to  $NL\rho\delta$ model. 

All these results can be easily fully understood remembering the
general relation between $(n,p)$ chemical potentials and symmetry energy
\cite{baranPR}:
\begin{equation}
\mu_n - \mu_p = 4 E_{sym}(\rho_B) \alpha .
\label{chemsym}
\end{equation}
For a fixed asymmetry $\alpha$ and baryon density $\rho_B$, 
one can directly
extract the equilibrium $\pi^-/\pi^+$ ratio, Eq.(\ref{pireq}). In the $NL$
model (no isovector interaction) we have only the weak Fermi contribution
to the symmetry energy, that largely increases, instead, 
when the $\rho$-coupling
is introduced. A further increase, but of smaller amplitude, is also
observed passing from $NL\rho$ to $NL\rho\delta$ models, see
Fig. \ref{fig1}.

Of course these equilibrium considerations do not hold for perturbative
kaon production, for which reabsorption
processes are neglected. 


\section{Results}
\label{chap4}
\begin{figure}[t]
\unitlength1cm
\begin{picture}(10.,9.0)
\put(0.0,0.0){\makebox{\epsfig{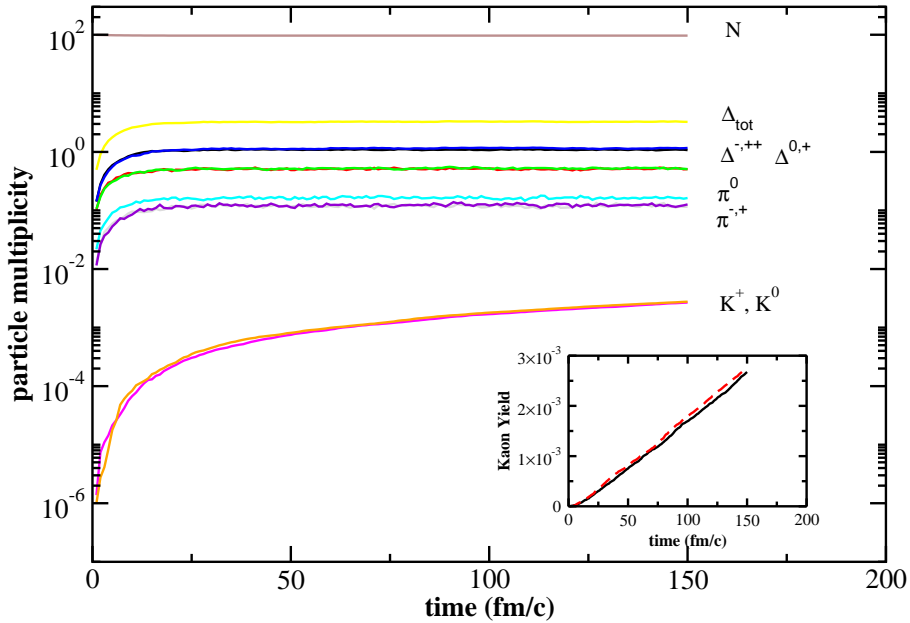}}}
\end{picture}

\caption{Time dependence of particle multiplicities (as indicated) 
for symmetric 
($\alpha=0$) hadronic matter at the same fixed conditions as in 
Fig. \protect\ref{fig2}. The inserted panel shows the temporal 
evolution of kaons in a linear scale.}
\label{fig3}
\end{figure}

Here we present the results for excited charge asymmetric matter 
within the effective field models of Sect. \ref{chap2}. The 
initial conditions of the 
matter in the box 
have been fixed to a baryon density of $\rho_{B}=2.5 \rho_0$ and a 
temperature of $T=60$ MeV, see again Fig. \ref{fig2}. We start from the 
simpler case of symmetric hadronic matter applying only the $NL$ model. 
Then we will discuss the isospin dependence (by varying the iso-vector part of 
the mean field) of 
particle production in the asymmetric case beginning with pions and $\Delta$
resonances. A similar discussion on strangeness production ($K^{+,0}$) 
concludes the section. 
 
Fig. \ref{fig3} shows the temporal evolution of nucleon, $\Delta$ resonance,
pion and kaon multiplicities for symmetric matter. Due to isospin symmetry, 
different isospin states for $\Delta$, $\pi$ and $K$ are almost equally 
populated. 
We see that the number of nucleons is almost constant throughout the time 
interval considered and just a tiny fraction of them (10$^{-2}$) goes into 
resonances since the first NN collisions. Some of these resonances further 
decay into one-pion channels with typical lifetimes of $\sim$1 fm/c and are 
subsequently 
reformed in $\pi$N collisions, according to isospin-dependent cross-sections 
taken from Ref. \cite{cugnon}. Therefore, the total 
numbers of $\Delta$ and pions grow up in a few fm/c and reach equilibrium 
values of $\sim$3.3 and $\sim$0.4 respectively within 30 fm/c, 
when formation and absorption mechanisms occur with almost the same 
rate. 

Kaons ($K^{0,+}$), on the other hand, do not equilibrate, because 
they cannot be 
reabsorbed after formation
and their number rises almost linearly with time, see the inserted panel
in Fig. \ref{fig3}. 
Their multiplicity is 
indeed well fitted
by a linear function of time of the form
\begin{displaymath}
N_{K}(t) \sim 1.9 \times 10^{-5} \: \: t - c
\end{displaymath}
\vskip -0.5cm
($c$ is a constant of the order of $10^{-4}$)
where the extremely small rate is due to the fact that in the conditions 
described we are
far from the kaon production threshold (1.56 GeV for NN collisions in 
free space)
 and the observed subthreshold production thus originate from extreme 
regions of 
phase space, i.e. from the tails of the momentum distributions of 
the particles involved.
Actually, kaons could undergo
inelastic collisions with other strange particles such as hyperons and thus
be converted into nonstrange particles. Anyway, studies performed at higher
energy densities \cite{bratko} have shown that, even taking such processes 
into 
account, strange particles  require a long time 
for
equilibration, of the order of several hundreds fm/c, and hence we can  
neglect here kaon reabsorption as a reasonable approximation,
 in particular for $K^{0,+}$. 

The above picture obviously does no longer hold for neutron-rich 
matter, where 
$\Delta^{-,0}$, $\pi^{-}$ and $K^{0}$ are favoured with respect to 
$\Delta^{+,++}$, $\pi^{+}$ and $K^{+}$. 
We will see in the following how the yields, and in particular
the ratios $\pi^{-}/\pi^{+}$ and  $K^{+}/K^{0}$, will be even sensitive
to the Lorentz structure of the isovector interaction.

\subsection{Isospin effects on pions and $\Delta$ resonances}

\begin{figure}[h]
\unitlength1cm
\begin{picture}(10.,9.0)
\put(0.0,0.0){\makebox{\epsfig{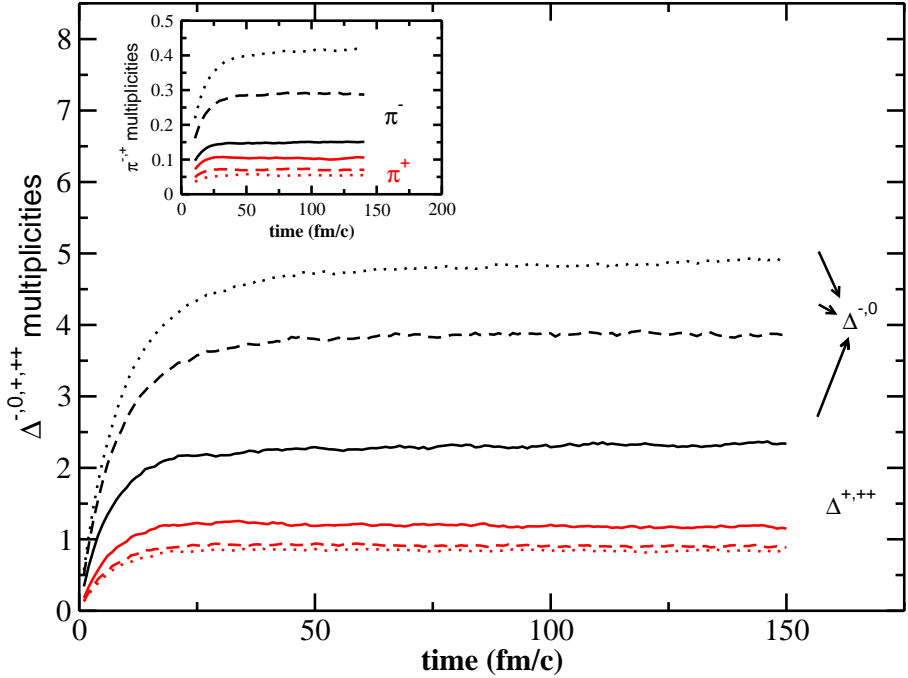}}}
\end{picture}
\caption{Time dependence of $\Delta$ resonances 
($\Delta^{-,0} \equiv \Delta^{-}+\Delta^{0}$), 
($\Delta^{+,++} \equiv \Delta^{+}+\Delta^{++}$) and pions 
(inserted panel), 
as indicated, for asymmetric ($\alpha=0.2$) hadronic matter at the same 
fixed conditions of density and temperature of Fig. \protect\ref{fig2}. 
Box calculations using different models for the mean field are shown:  
($NL$, solid lines) including only the iso-scalar part 
($\sigma,~\omega$ mesons), 
($NL\rho$, dashed lines) accounting for the iso-vector, 
vector $\rho$ field  and 
($NL\rho\delta$, dotted lines) taking into account both 
iso-vector $\rho$ and 
$\delta$ fields.
}
\label{fig4}
\end{figure}

Fig. \ref{fig4} shows the effect of the isovector channels on $\Delta$ 
and $\pi$ multiplicities, for the same total density and temperature
starting conditions but with a fixed asymmetry $\alpha=0.2~(N/Z=1.5)$. 
In the simple $NL$ model the relative yields of 
positively 
and negatively charged pions and/or the corresponding resonances, just follow 
from the different numbers of {\it nn} and 
{\it pp} pairs and, consequently, of their inelastic collisions: 
$nn\rightarrow p\Delta^{-}$,
$\Delta^{-} \rightarrow n \pi^{-}$ and $pp\rightarrow n\Delta^{++}$,
$\Delta^{++} \rightarrow p \pi^{+}$. Clearly the $\pi^{-}$ 
multiplicity 
increases with asymmetry parameter, i.e. as the matter becomes more and 
more neutron-rich
while
the number of $\pi^{+}$ accordingly decreases  with respect to the symmetric 
case where we have $\pi^{-}=\pi^{+}$ yields. 
The further inclusion of a $\rho$ meson ($NL\rho$ model), 
stresses such opposite 
trend in the charged particle production, with the
result that the former isospin states become even more populated in the case of
NL$\rho$ model, while the latter are less populated
 (solid to dashed lines in Fig. \ref{fig4}). 
Therefore, the number of $\pi^{-}$s, which originate 
from $\Delta^{-,0}$ decay, also increases, while $\pi^{+}$s  decrease
(inserted panel in Fig. \ref{fig4}).

Finally on the basis of the argument discussed in the previous section, one 
expects a further enhancement of 
the effect by the inclusion of the isovector scalar $\delta$ meson,
as confirmed by the results presented in the Figure.

\begin{figure}[t]
\unitlength1cm
\begin{picture}(10.,9.0)
\put(0.0,0.0){\makebox{\epsfig{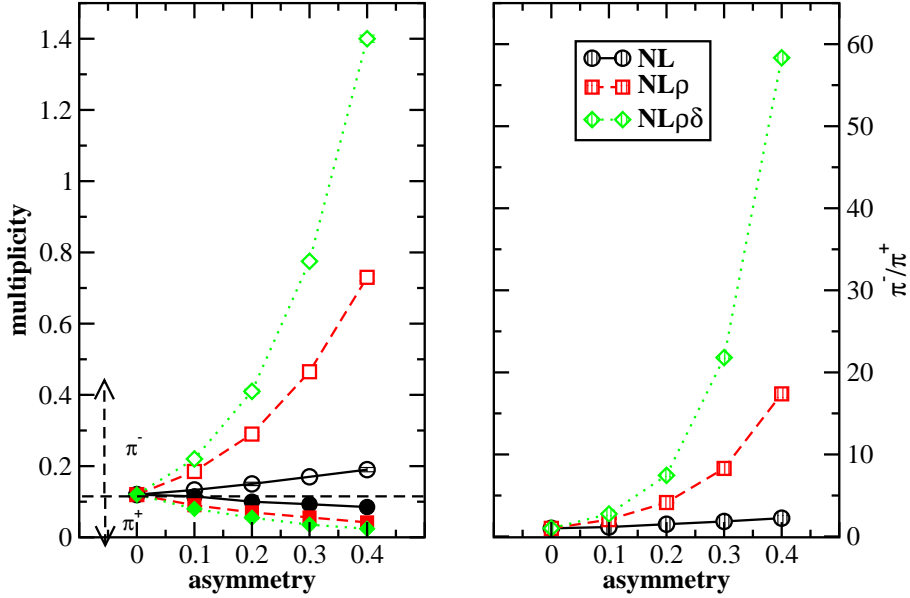}}}
\end{picture}
\caption{Charged pion multiplicities (left) and the $\pi^{-}/\pi^{+}$ ratio 
(right) as a function of the asymmetry parameter $\alpha$ 
at the same conditions of Fig. \protect\ref{fig2}. Statistical errors
are inside the data points.
}
\label{fig5}
\end{figure}
In Fig. \ref{fig5} we show the asymmetry dependence of the charged pion 
multiplicities.
When chemical equilibrium is reached, the $\pi^{-}
/ \pi^{+}$ ratio depends just on the difference between neutron and
proton chemical potentials and on the temperature $T$. 
Indeed, if we look at this 
ratio as a function of the $\alpha$  parameter (right panel) we clearly see 
the behaviour already noticed from Fig. \ref{fig4}: 
the $\pi^{-}/\pi^{+}$ ratio increases when passing 
from $NL$ 
to the $NL\rho$ and then to the $NL\rho\delta$ model.
As explained previously, the $\pi^{-}/ \pi^{+}$ ratio should increase
by a factor $exp~( 4f_\rho(\rho_{Bn}-\rho_{Bp})/T)$  when passing from 
NL to $NL\rho$ model.  For instance, at the asymmetry $\alpha = 0.2$, 
this factor equals
$3.6$, that approximately corresponds to the enhancement observed in 
the Figure. 

At the same asymmetry, we get for the effective neutron and 
proton chemical
potentials: $\mu^*_n - \mu^*_p~=~34~MeV$ for $NL$ and $NL\rho$ models and 
$-16~MeV$ for the $NL\rho\delta$ model, due to the nucleon Dirac mass 
splitting Eq.(\ref{emstar}).
Hence, considering the change in the difference of effective chemical
potentials and the value of the constant $f_{\rho}$ (see Tab.I), 
we see that 
the difference $\mu_n - \mu_p$, to which the  $\pi^{-}$/$\pi^{+}$ ratio
is related, changes   
when going from $NL\rho$ to $NL\rho\delta$ model and 
we expect an increase of this ratio by a factor $1.88$. 
This is also nicely confirmed by our results.

\subsubsection{ Isospin effects on pion reabsorption }

Since in nuclear collisions equilibrium conditions 
are likely not reached, it is interesting
to calculate just the probability of emitting pions in the other extreme case,
i.e. excluding the pion reabsorption processes.
Now the  observed $\pi^{-}$/$\pi^{+}$ ratio should reflect directly 
the ratio between the cross sections for  $\pi^{-}$ and $\pi^{+}$
production.
  
\begin{figure}[t]
\unitlength1cm
\begin{picture}(10.,7.0)
\put(2.0,0.0){\makebox{\epsfig{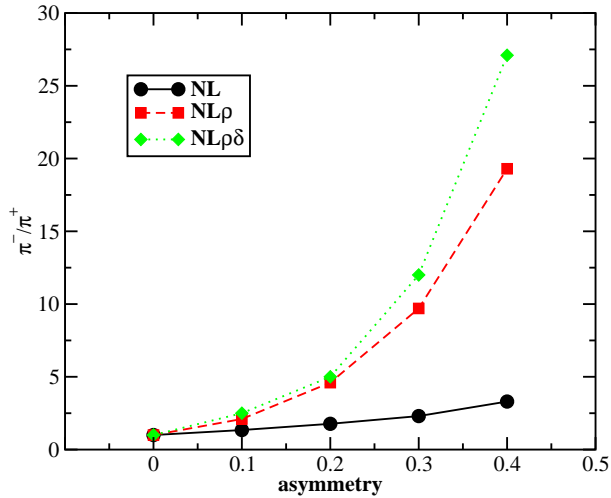}}}
\end{picture}
\caption{Asymmetry dependence of the $\pi^{-}/\pi^{+}$ ratio for the 
same models as in Fig. \ref{fig2}, but without the inclusion of 
reabsorption.
}
\label{fig6}
\end{figure}

As shown in Fig. \ref{fig6}, the
$\pi^{-}/\pi^{+}$ ratio rises both in the $NL\rho$ (with respect to the $NL$) 
model 
and 
in the $NL\rho\delta$ (with respect to the $NL\rho$) model, for any value 
of the 
asymmetry parameter. In fact, as discussed in the previous sections,
the addition of 
each isovector meson ($\rho$ and $\delta$) favours $\pi^{-}$ over $\pi^{+}$ 
production, as a result of the higher (lower) energy
available in the ``parent'' $nn$ ($pp$) collisions.

We notice that the difference of $\pi^{-}/\pi^{+}$ ratio 
observed among the different models is less
pronounced with respect to the results observed when equilibrium
is reached (see Fig. \ref{fig5}).  In particular, the results obtained with
$NL\rho$ or $NL\rho\delta$ are quite close.
Moreover, in a nucleus-nucleus collision, pions interact with the 
surrounding matter
all along the reaction path. Hence the observed final pion multiplicity 
is not clearly connected to the very first stage of the collision and to the
high density behaviour of the symmetry energy.   
In the next section we will discuss kaon production, for which we expect a
larger sensitivity to the models used, since in the conditions considered here 
we are close to the production threshold.
Moreover, kaons only loosely interact with the nuclear medium and consequently 
they should carry a better signal of the symmetry energy behaviour at large 
density.

\subsection{Isospin effects on kaon production}
In the case of kaon production the isospin states to 
be considered are 
$K^{0}$ and $K^{+}$, which can be produced through a large variety of channels
with two or three particles in the final state: 
$BB \longrightarrow BYK,~B\pi \longrightarrow YK$ ($B,Y,K$ stand for 
non-strange baryon, 
hyperon
and kaon, respectively). 
While, for the baryon channels, a proton-proton collision can produce 
$\pi^{+}$ but not $\pi^{-}$ and 
a neutron-neutron collision, conversely, only $\pi^{-}$ but not $\pi^{+}$, 
in the 
case of kaons isospin conservation allows $K^{0}$ and $K^{+}$ to be formed 
from both initial {\it pp} and {\it nn} scattering processes. However, due to 
isospin coefficients, $K^{0}$ mainly come from {\it nn} inelastic channels and 
$K^{+}$ from {\it pp} channels. We thus expect their relative yields to 
change with
the asymmetry parameter $\alpha$, similarly to pions without 
reabsorption, but with a larger sensitivity since kaons are produced 
close to the threshold.  
Moreover, from the previous discussion on pion production,
also the pionic channels will contribute to isospin effects on
kaon production. 

The $BB \longrightarrow BYK,~B\pi \longrightarrow YK$ channels can be 
schematically divided as follows:
\begin{enumerate}
\item $N N \longrightarrow BYK$
\item $N \Delta \longrightarrow BYK$
\item $\Delta \Delta \longrightarrow BYK$
\item $\pi N \longrightarrow YK$
\item $\pi \Delta \longrightarrow YK$
\end{enumerate}
where $N$  stands for a nucleon and $B$ for a non-strange baryon.

At low energy densities, such as those achieved in the present calculations 
and/or 
in heavy ion collisions at energies below 1.5 AGeV, the pionic channels 
$\pi N$, (iv), give
the main contribution \cite{fuchspion}, together with the $N \Delta$ 
ones, (ii).
 $\pi\Delta$ 
and $NN$ collisions contribute just for a few percent to the total 
$K^{+}$ and/or
$K^{0}$ yield, while the $\Delta \Delta$ channel is negligible. 

The advantage of a cascade calculation in a box is that we can cleanly 
evaluate the relative weight of the various kaon production channels
in the interacting matter, i.e. vs. the different choices of the
hadron self-energies. Our results are reported in Table \ref{table2}.
We see that for 
all the iso-vector models discussed here we are in agreement with
previous estimations \cite{fuchspion}. We note also, in the case of
asymmetric matter, the different behavior of the important $\pi N$
channel when we add the isovector mesons: the weight is decreasing
for the $K^+$ production while it is increasing for the $K^0$.
This is very promising for our search for fine relativistic effects
of the isovector interaction on kaon production. 

\begin{table}[t]
\begin{center}
\caption{\label{table2} 
Relative weight of various $K^{0,+}$ production channels for symmetric 
($\alpha=0$) and asymmetric ($\alpha=0.2$) matter. The relative width 
for the $\Delta\Delta$ channels is given in units of $10^{-4}$. 
Calculations with different symmetry energies are shown too.
}
\vskip 0.5cm
\begin{tabular}{c|l|ccccc|ccccc}
\hline\hline
     &   &   &   & $K^{+}$  &
        &   &   &   &  $K^{0}$ & \\
\hline\hline
$\alpha$  &  & $NN$ & $N\Delta$ & $\Delta\Delta$ & $\pi N$ & $\pi \Delta$
     & $NN$ & $N\Delta$ & $\Delta\Delta$ & $\pi N$ & $\pi \Delta$ \\ 
\hline
    0   &  $NL$             & 0.032    &   0.232  & $1$ & 0.643 & 0.093 
                    & 0.030    &   0.221  & $3$ & 0.665 & 0.083 \\ 
    0.2 &  $NL$             & 0.033     &   0.206  & $1$ & 0.670 & 0.091 
                    & 0.029    &   0.228  & $2$ & 0.651 & 0.085 \\ 
    $''$         &  $NL\rho$         & 0.030    &   0.227 & $< 1$       & 0.633 & 0.119    
                    & 0.026    &   0.231  & $3$ & 0.662 & 0.072 \\ 
    $''$         &  $NL\rho\delta$   & 0.033    &   0.236 & $2$ & 0.591 & 0.131   
                    & 0.024    &   0.192 & $3$ & 0.714 & 0.070 \\ 
\hline\hline
\end{tabular}
\end{center}
\vskip 0.5cm
\end{table}
\begin{table}[t]
\begin{center}
\caption{\label{table3} 
Production rate of $K^{+,0}$ (in units of $10^{-5}$) for asymmetric 
($\alpha=0.2$) hadronic matter for the $NL,~NL\rho,~NL\rho\delta$ models.
 }
\vskip 0.5cm
\begin{tabular}{|l|c|c|c|c|c|}
\hline\hline 
       & $\frac{dN(K^{+})}{dt}$ ($\times 10^{-5})$  & $\frac{dN(K^{0})}{dt}$ ($\times 10^{-5})$\\ 
\hline\hline
   $NL$             &     1.66        &   2.18   \\ 
\hline
   $NL\rho$         &     1.17        &   2.91    \\ 
\hline
   $NL\rho\delta$   &     1.06       &    3.92     \\ 
\hline\hline
\end{tabular}
\end{center}
\vskip 0.5cm
\end{table}
\begin{figure}[t]
\unitlength1cm
\begin{picture}(10.,9.0)
\put(0.0,0.0){\makebox{\epsfig{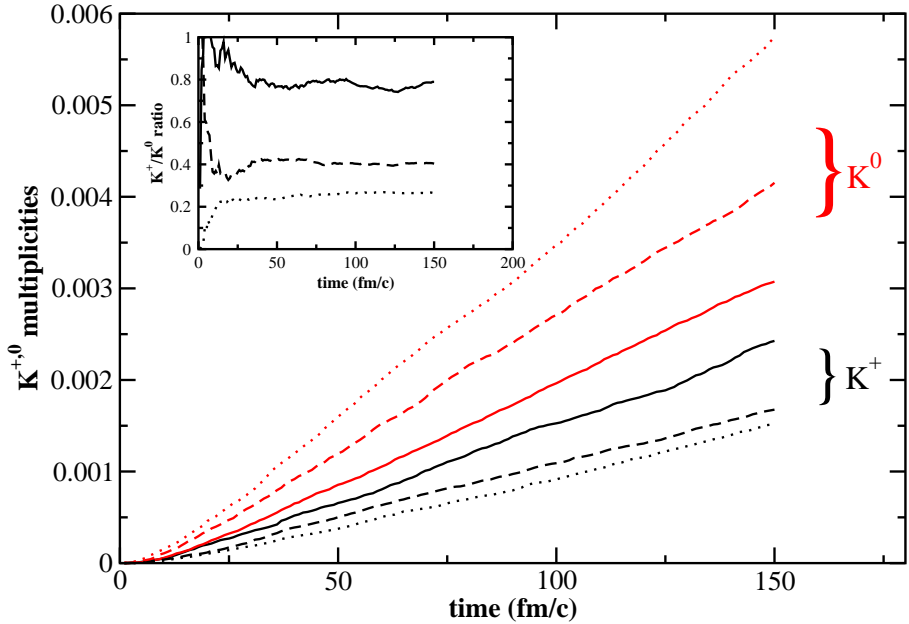}}}
\end{picture}
\caption{
Temporal evolution of $K^{+,0}$ yields and their ratio (in the inserted panel) 
for asymmetric ($\alpha=0.2$) hadronic matter for the same iso-vector 
models as 
in Fig. \protect\ref{fig2}.
}
\label{fig7}
\end{figure}
\begin{figure}[t]
\unitlength1cm
\begin{picture}(10.,9.0)
\put(0.0,0.0){\makebox{\epsfig{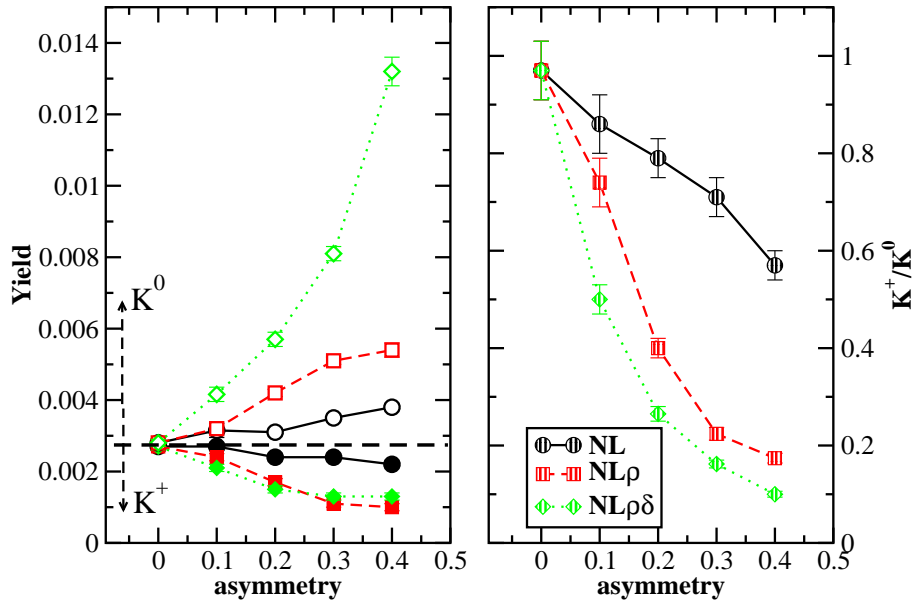}}}
\end{picture}
\caption{
Asymmetry dependence of the $K^{+,0}$ yields (left) and of the 
$K^{+}/K^{0}$ ratio 
(right) for the different iso-vector models as in Fig. 
\protect\ref{fig2}. Statistical error bars are also shown,
when larger than the data points.
}
\label{fig8}
\end{figure}

Fig. \ref{fig7} shows the 
temporal evolution of kaon multiplicities for asymmetric 
($\alpha=0.2$ or N/Z=1.5 corresponding to the asymmetry 
of ${}^{197}Au$) hadronic matter and their ratio for the $NL$, $NL\rho$ and 
$NL\rho\delta$ models. First of all, we observe an almost linear time 
dependence of the kaon yields as already discussed (see Fig. \ref{fig2}). 
The isospin effect on the kaon yields essentially turns out in a change of 
the slope, 
i.e. of the production rate, see Table \ref{table3}.
In particular, for $K^{0}$ ($K^{+}$) the formation rate increases (decreases) 
as the symmetry energy becomes stiffer, i.e. passing from the $NL\rho$ to the 
$NL\rho\delta$ case.

The whole picture is summarized in Fig. \ref{fig8}, where the kaon yields and 
their ratio are displayed as a function of the asymmetry parameter $\alpha$. 
The influence of the different iso-vector models obviously increases with 
 $\alpha$ for the $K^{0}$ yield, but less for the 
$K^{0}/K^{+}$ ratio.
In fact, as already shown in Fig. \ref{fig7} and Table \ref{table3} for a 
fixed asymmetry, 
the  $K^{+}$ yield, which decreases of about $25\%$ and $10\%$ in NL$\rho$
and NL$\rho\delta$ models respectively, presents an isovector meson 
effect slightly 
weaker 
than the $K^{0}$ 
yield, which exhibits increasing variations of about $35\%$ for both 
 $NL\rho$ 
and $NL\rho\delta$ models.

Generally the isospin effects on kaon yields originate from at least two 
different mechanisms: the isospin dependencies of the $\pi^{\pm}$ 
yields and of the threshold energies. We will now discuss in detail these 
two effects.

Since kaon production from pionic channels (which give the main contribution)
is competitive with pion reabsorption, it occurs before secondary 
$\pi N \longrightarrow \Delta$ processes set in. 
Therefore, 
$K^{0}$ and $K^{+}$ kaon yields are mainly affected by the $\pi^{-}$ and 
$\pi^{+}$ 
abundances before reabsorption, see Fig. \ref{fig6}. In particular, as 
$K^{0}$ ($K^{+}$) mainly come 
from $\pi^{-} N$ ($\pi^{+} N$) processes, the enhancement (reduction) of the 
$\pi^{-}$ ($\pi^{+}$) multiplicities in the $NL\rho\delta$ model is partially 
responsible for a corresponding increase (decrease) of the  $K^{0}$ ($K^{+}$) 
yields. 

The isospin effect on the kaon production threshold is more complicated,  
due to the many isospin channels that may affect the threshold 
(and hence the yields) 
differently. However, contributions in opposite directions originating from 
different channels 
do not exactly cancel each other. 
In order to clarify the observed 
influences of the models on the yields in a transparent way, it is appropriate 
to focus on some demonstrative channels. Such examples  
are the isospin channels $n\pi^{-} \longrightarrow \Sigma^{-} K^{0}$ and 
$p\pi^{+} \longrightarrow \Sigma^{+} K^{+}$ for $K^{0,+}$ production, 
respectively.

$\pi N$ channels offer the advantage of two particles in the final state
(instead of the three that we have in the $BB$ case); moreover, since also 
kaons (and not only pions)
are, in the present calculations, unaffected by nuclear mean field, 
Eqs. (\ref{smod}-\ref{sin})
are simplified ($\Sigma_{i2}=\Sigma_{i \pi}=\Sigma_{i4}=\Sigma_{ik}=0$).  

According to Eq. (\ref{smod}) the corresponding threshold conditions read 
\begin{eqnarray}
s_{in}(n \pi^{-}) \geq (M_{\Sigma^-} + M_{K^0} + \Sigma_{s \Sigma^-}
 + \Sigma^{0}_{\Sigma^-})^2 + {\bf \Sigma}_{\Sigma^{-}}^{2} 
\label{smod2}\\
s_{in}(p \pi^{+}) \geq (M_{\Sigma^+} + M_{K^+} + \Sigma_{s \Sigma^+}
 + \Sigma^{0}_{\Sigma^+})^2 + {\bf \Sigma}_{\Sigma^{+}}^{2}
\label{smod3}
\quad ,
\end{eqnarray}
where $M_{\Sigma^-} + M_{K^0} = 1695 GeV$ and
 $M_{\Sigma^+} + M_{K^+} = 1683 GeV$.

As discussed in the previous section, for collisions involving neutrons 
(protons)
$s_{in}$ increases (decreases)
when going from $NL$ to $NL\rho$ and then to $NL\rho\delta$ model. 
Therefore, the available energy for $K^{0}$ production increases as the 
symmetry energy becomes stiffer, leading to an enhanced $K^{0}$ yield. 
Obviously the trend for the 
isospin behavior of the $K^{+}$ yield will be opposite.

\section{Conclusions}
\label{chap5}

We have investigated the high density behavior of the 
iso-vector of the nuclear interaction, which is still poorly known 
experimentally, controversially predicted by theory, but of great 
interest in extreme nuclear systems. Here we discuss its dynamical
effects on particle production 
for an idealized system, the asymmetric hadronic 
matter. In this respect relativistic cascade calculations have
been performed in a box with periodic boundary conditions. The advantage
is that in this way we can cleanly follow all the contributions
of the various isospin channels to the production and reabsorption,
and the corresponding dependence on the effective field structure
of the isovector interaction.

The paper represents a detailed study, in a relativistic frame, of
the production/absorption mechanisms of energetic particles (pions
and kaons) in excited nuclear matter at high isospin and baryon density
 and temperature. These are transient conditions in realistic
charge asymmetric heavy ion collisions at intermediate energies.
The yields are largely influenced by the isospin dependence of the nucleon 
self-energies. In particular, a given isospin state of the produced particles 
is differently populated varying the high density behaviour of 
the symmetry energy.

It has been found that, after $\approx 30$ fm/c, chemical equilibrium 
is reached.
The $\pi^{-}/\pi^{+}$  ratio depends on the model considered, 
according to the difference
between neutron and proton chemical potentials. 
Excluding the reabsorption process, in order to measure just the pion
emission probability in situations when equilibrium is not necessarily 
reached, 
we find a less pronounced dependence of 
the $\pi^{-}/\pi^{+}$ ratio  on the symmetry energy behaviour. 
This seems somehow confirmed by theoretical transport calculations 
on Au+Au collisions \cite{gait04,bao02}.

The isovector meson effect on kaon rates  
appears larger than on pions (excluding reabsorption) because 
kaons are produced close to the threshold. 
It should be noticed that 
low energy kaons do not suffer from secondary reabsorption 
effects neither 
in infinite hadronic matter nor in collisions of finite nuclei due 
to strangeness 
conservation.  Hence, in a nuclear collision, they are expected 
to better signal the behaviour of
the symmetry energy during the first stage of the collision, when they are 
formed.
As an important result, 
 the $K^{0}$ 
and $K^{+}$ yields are oppositely affected leading to a crucial isospin 
dependence of the $K^{+}/K^{0}$ ratio, which one would expect to observe 
in heavy ion collisions at subthreshold energies. 
Such an investigation is being 
performed and will be discussed in a future work.

In conclusion the results presented here appear very promising for 
the possibility
of extracting information from terrestrial laboratories on the
Lorentz structure of the isovector nuclear interaction in a medium
at densities of astrophysical interest. The use of radioactive beams
at relativistic energies would be extremely important.




\begin{thebibliography}{99}
\bibitem{sigma} 
This choice will not affect the results discussed here, when a chemical 
equilibrium is reached. Moreover we note that even for an ``open'' system,
such as a realistic heavy ion collision, an in-medium (reduced) parametrization
will just lead to a rescaling of the particle, resonance multiplicities,
without much changing the relative yields we are focussing on.
\end{thebibliography}
\end{document}